\documentclass[letterpaper, 10 pt, conference]{ieeeconf}  % Comment this line out if you need a4paper

\IEEEoverridecommandlockouts                              % This command is only needed if 
\usepackage[utf8]{inputenc}

\usepackage{graphics} % for pdf, bitmapped graphics files
\usepackage{bm}
\usepackage[dvipsnames]{xcolor}
\usepackage{draftwatermark}
\usepackage{framed}
\usepackage{url}

\usepackage[backend=bibtex,style=ieee,doi=true,isbn=false,url=false,eprint=false,maxnames=3,citestyle=numeric-comp,giveninits=true]{biblatex}

\bibliography{references.bib}

\SetWatermarkText{accepted paper}
\SetWatermarkScale{0.4}
\SetWatermarkColor[gray]{0.9}
\definecolor{light-gray}{gray}{0.95}

\renewcommand{\vec}[1]{\bm{#1}}

\usepackage{fancyhdr}
\title{\LARGE \bf
HEAR to remove pops and drifts: the high-variance electrode artifact removal (HEAR) algorithm
}

\author{Reinmar J. Kobler$^{1}$, Andreea I. Sburlea$^{1}$, Valeria Mondini$^{2}$ and Gernot R. Müller-Putz$^{1}$% <-this % stops a space
\thanks{$^{1}$Institute of Neural Engineering, Graz University of Technology, Graz, Austria
        {\tt\small \{reinmar.kobler, gernot.mueller\}@tugraz.at}}%
\thanks{$^{2}$Department of Electrical, Electronic and Information Engineering, \mbox{University of Bologna}, Bologna, Italy}%
}

\begin{document}

\maketitle
\thispagestyle{fancy}
\pagestyle{empty}

%%%%%%%%%%%%%%%%%%%%%%%%%%%%%%%%%%%%%%%%%%%%%%%%%%%%%%%%%%%%%%%%%%%%%%%%%%%%%%%%
\begin{abstract}
A high fraction of artifact-free signals is highly desirable in functional neuroimaging and brain-computer interfacing (BCI).
We present the high-variance electrode artifact removal (HEAR) algorithm to remove transient electrode pop and drift (PD) artifacts from electroencephalographic (EEG) signals.
Transient PD artifacts reflect impedance variations at the electrode scalp interface that are caused by ion concentration changes.
HEAR and its online version (oHEAR) are open-source and publicly available.
Both outperformed state of the art offline and online transient, high-variance artifact correction algorithms for simulated EEG signals.
(o)HEAR attenuated PD artifacts by approx. 25\,dB, and at the same time maintained a high SNR during PD artifact-free periods.
For real-world EEG data, (o)HEAR reduced the fraction of outlier trials by half and maintained the waveform of a movement related cortical potential during a center-out reaching task.
In the case of BCI training, using oHEAR can improve the reliability of the feedback a user receives through reducing a potential negative impact of PD artifacts.
\end{abstract}

%%%%%%%%%%%%%%%%%%%%%%%%%%%%%%%%%%%%%%%%%%%%%%%%%%%%%%%%%%%%%%%%%%%%%%%%%%%%%%%%
\section{INTRODUCTION}

Electroencephalography (EEG) is a widespread non-invasive functional neuroimaging technique to study electrophysiological activity in the brains of humans \cite{Schomer2011}.
EEG signals are recorded by sampling the voltage between electrodes and a reference electrode at the scalp.
During this process, not only brain signals, but also other physiological and non-physiological signals are captured.
Other physiological signals such as electromyographic (EMG), electrooculographic (EOG) and electrocardiographic (ECG) signals as well as non-physiological signals such as power-line noise or  impedance variations at the electrode scalp interface are typically undesired and classified as artifacts.
The impedance variations at the electrode scalp interface manifest as pops and drifts (PD) in the recorded EEG signals.
In EEG-based brain-computer interfacing (BCI), transient, high-variance PD artifacts can temporarily deteriorate the control signal and, thereby, impede closed-loop control.
It is, therefore, desirable to remove or at least detect PD artifacts.

%The impedance variations at the electrode scalp interface can be caused by various reasons.
Key to reduce impedance variations is to properly attach the electrodes to the scalp \cite{Anderer1999}.
To date, the best long-term stability is achieved with sintered Ag/AgCl electrodes and a salty ($Cl^-$) electrolyte \cite{Tallgren2005}.
In this case, the impedance typically stabilizes after approximately 20 to 30 minutes.
Nonetheless, transient PD artifacts can arise due to fluctuations of $Cl^-$ concentration in the electrolyte.
The change in $Cl^-$ concentration can be caused by sweating, drying electrolyte, or by liquefaction of a crust of dried electrolyte that covers the electrode.

The variance of transient PD artifacts is typically much higher than the variance of ongoing brain activity.
Electrode pops can be described with a step and a subsequent exponential decay.
As such, they have a broad-band spectrum with highest spectral power in the lower frequencies.
Transient electrode drifts can be described with band-limited, low-frequency -- typically $\le 0.25\,Hz$ \cite{Hammond2001} -- noise.
As a consequence, BCIs decoding slow processes such as movement related cortical potentials (MRCP) \cite{Shibasaki1980} are most prone to PD artifacts.

In recent years, many automatic artifact cleaning methods have been introduced \cite{Uriguen2015}.
Some are particularly suitable to remove transient, high-variance artifacts.
Offline, it is common practice to detect transient artifacts within trials by visual inspection. 
Alternatively, outlier trials can be detected automatically using thresholding or high-order statistics \cite{Delorme2001}.
In case of PD artifacts, contaminated trials are rejected or the signals of the affected electrodes interpolated \cite{Perrin1989}.

Kothe and Jung introduced the artifact subspace reconstruction (ASR) algorithm \cite{Kothe2014}.
ASR is a variance-based method and has been shown to improve the quality of independent component analysis decompositions \cite{Pion2018}.
ASR applies principal component analysis (PCA) in a sliding-windowed approach.
The variance of each principal component (PC) is compared to a threshold which is derived from calibration data.
Since the orientation of the PCs can change in each new data window, the thresholds are projected to the new PCs.
A PC is removed, if its variance is larger than the variance during the calibration data multiplied by a cutoff parameter $k$.
The corrected EEG is computed by back-projecting all clean PCs to the original electrode space.

Another suitable algorithm is robust PCA (RPCA).
RPCA was originally designed to partition surveillance videos into transient and stationary segments \cite{Candes2011}.
A video represented as an $ N_{features} \times N_{samples}$ matrix $\vec{X}$ would be decomposed into the sum of a sparse matrix $\vec{S}$ and a low-rank matrix $\vec{L}$.
$\vec{S}$ and $\vec{L}$ can be estimated by solving the optimization problem

\begin{equation}
min ~ || \mathbf{L} ||_* + \frac{\lambda_0}{\sqrt{N_{samples}}} || \mathbf{S} ||_1 ~~s.t.~~ \mathbf{X} =  \mathbf{L} + \mathbf{S}
\label{eq:rpca:definition}
\end{equation}

with $\lambda_0$ being a regularization parameter that trades-off between the nuclear norm of $\vec{L}$ and the L1 norm of $\vec{S}$.
In the context of BCIs, RPCA has been used to reduce session-to-session \cite{Jao2015} and trial-to-trial \cite{Jao2018} variability, and recently also transient, high-variance artifacts \cite{Kobler2018}.
RPCA is only applicable offline, since it decomposes the entire data at once into corrected EEG signals $\vec{L}$ and transient artifacts $\vec{S}$.

The reported results seem to support the application of ASR and RPCA to remove transient, high-variance artifacts. 
However, literature lacks a thorough investigation of their performance - specifically with regard to PD artifacts.
In this paper, we evaluated the performance of ASR and RPCA on one simulated and one real-world EEG dataset.
We compared the results of ASR and RPCA with the high-variance electrode artifact removal (HEAR) algorithm which is introduced in the next section.
Unlike ASR and RPCA, HEAR was specifically designed to remove PD artifacts.
We hypothesized that all algorithms could improve the SNR during PD artifacts, and that HEAR would be superior to ASR and RPCA since it utilizes the structure of PD artifacts.

\section{METHODS}

\subsection{High-variance Electrode Artifact Removal (HEAR)}
The algorithm is based on two assumptions which are typically met by 
PD artifacts.
First, the variance of each electrode signal can be used to detect periods of PDs.
Second, PDs typically appear at single or few electrodes.
For a sufficient spatial resolution, the neighboring electrode signals can be temporarily used to estimate the signal of the contaminated electrode.

The detection of PD artifacts based on the electrode variance requires a reference variance $ \mu_s^{2\,(i)} $ for each electrode $i$.
The $N_{electrodes} ~ \times ~ 1 $ vector $ \vec{\mu}_s^2 $ is computed as the average variance during calibration (e.g., resting) data with no or few PD artifacts.
The HEAR algorithm uses an exponential smoothing filter to estimate the instantaneous variances $ \vec{s}^2[n] $ at time $n$ as

\begin{equation}
    \vec{s}^2[n] = \lambda \vec{s}^2[n-1] + (1 - \lambda) \vec{x}^2[n]
    \label{eq:hear:filter}
\end{equation}

with the smoothing factor $ \lambda $ defined as

\begin{equation}
    \lambda = (1-p)^{\frac{1}{t_{est} \cdot f_s}}
\end{equation}

so that the time window $t_{est}$ receives $p$ percent of the weights at the sampling rate $f_s$.
In this paper, we used \mbox{$p = 0.9$}.

Once the reference variances are computed, the HEAR algorithm can be applied to correct new samples $\vec{x}[n]$.
The correction process is implemented in three steps.
First, equation (\ref{eq:hear:filter}) is applied to update the estimate of the electrode variances $ \vec{s}^2[n] $.
Second, the probability that $ s^{(i)}[n] $ was caused by an artifact $p_{art}( s^{(i)}[n]) = p(art \le s^{(i)}[n]) $ is derived from a normal distribution

\begin{equation}
     p_{art}(s^{(i)}[n]) \sim \mathcal{N}(\phi\cdot \mu_s^{(i)},\, \xi^2 \cdot \mu_s^{2\,(i)})    
\end{equation}

with $\phi$ and $\xi$ being hyper-parameters to scale the mean and variance.
%In detail, $p_{art}( s^{(i)}[n]) = p(art \le s^{(i)}[n]) $ refers to the cumulative distribution function of the normal distribution.
The probabilities for all electrodes are combined to a single diagonal matrix $\vec{P}[n]$

\begin{equation}
    \vec{P}[n] = diag(p_{art}(s^{(1)}[n]),\,...,\,p_{art}(s^{(N_{channels})}[n]))
\end{equation}

Third, the corrected signal $\vec{x}_c[n]$ is computed via linear interpolation. $\vec{P}[n]$ is applied to weigh the amount of linear interpolation so that

\begin{equation}
    \vec{x}_c[n] = \vec{P}[n] \,\vec{D}\, \vec{x}[n] + (\vec{I} - \vec{P}[n])\,  \vec{x}[n]
    \label{eq:hear:correction}
\end{equation}
with $\vec{D}$ containing the relative distances.
The relative distances are the inverse Euclidean norm between the 3D position of the target electrode and its $k = 4$ nearest neighbors.
They are normalized so that the rows of $\vec{D}$ sum to 1.

In case of an offline analysis, the filter in (\ref{eq:hear:filter}) can be applied bidirectionally during the calibration and correction procedures.
In this paper, we refer to the bidirectionally filtered version as HEAR and the causally filtered version as online HEAR (oHEAR).
A reference implementation of (o)HEAR is publicly available at
%{\tt\footnotesize https://bci.tugraz.at/research/software/\#c218405}
\url{https://bci.tugraz.at/research/software/#c218405}

HEAR and oHEAR depend on three hyper-parameters $ \Theta = \{t_{test}, \phi, \xi\}$.
All parameters are intuitive to interpret.
The variance estimation duration $t_{est}$ trades-off between smoothness of the estimate and responsiveness to fast events such as pops.
The scaling factors of the artifact distribution $\{\phi, \xi\}$ define how often the reference variance has to be exceeded so that the artifact probability is 50\% ($\phi$), and how quickly the distribution increases ($\xi$).
Hence, $\phi$ and $\xi$ control the sensitivity of the algorithm.

In this paper, we varied the sensitivity of all correction algorithms by evaluating different hyper-parameter configurations.
ASR was evaluated for the cut-off parameter $k = \{20,\,40,\,80\} $ according to the recommendations in \cite{Chang2018}, and the default window size ($0.5$\,s).
Based on \cite{Kobler2018,Jao2018}, RPCA was evaluated for the regularization parameter $\lambda_0 = \{1.0,\,1.5,\,2.0\} $.
We controlled the sensitivity of (o)HEAR by setting $\phi = \{2,\,3,\,4\}$.
Using real data of pilot studies, we set $t_{est} = 0.25\,$, $\xi = 1 $.

% Properties:

% \begin{itemize}
% 	\item Channel locations required to compute $D$
% 	\item Online and offline possible. Offline exp-smoothing can be applied bidirectionally (zero-phase).
% \end{itemize}

We validated the performance of (o)HEAR, RPCA and ASR by applying them to one dataset of simulated EEG and one real-world EEG dataset.

\subsection{Simulated EEG dataset}

We generated a simulated dataset specifically for this study with the simulated event-related EEG activity (SEREEGA) toolbox \cite{Krol2018} and Matlab 2015b (Mathworks Inc., USA).
In detail, we simulated EEG signals at 64 electrodes as linear mixtures of sources on the cortical surface of the ICBM-NY head model template \cite{Huang2016} and the EEG electrodes themselves.
The 64 electrodes were placed according to the extended 10/20 system.
The simulated sources comprised oscillatory brain activity, an MRCP, and noise sources at the electrodes.
The noise sources were modeled as stationary white measurement noise and transient PD artifacts.

We simulated 15 participants.
For each participant, we used the same head model template, while the source locations and signals were independent and identically distributed (iid).
If not explicitly stated, a uniform distribution within a given range was used.
For each participant, we simulated two experimental tasks.
During the first task (rest) the oscillatory brain and stationary electrode noise sources were active.
We simulated 12 trials, each lasting 15\,s.
During the second task (reach) all sources were active.
We simulated 60 trials, each lasting 15\,s.

We modeled the oscillatory brain activity with 40 pink and 40 brown noise sources with amplitudes of $37.5\,\mu V$ and $75\,\mu V$ respectively.
The locations of the 80 sources were picked randomly from 74382 available locations on the template head model.

The location of the MRCP source was picked randomly in a 10\,mm radius around the coordinate $ [-25,\, 0,\, 80] $ \,mm, which corresponded to the medial end of the pre-central gyrus of the left hemisphere.
The waveform of the MRCP was modeled with radial basis function kernels so that the waveform started with a slow negative deflection 7\,s after the start of each trial, intensified abruptly after 700\,ms, peaked with an amplitude of -120 $\mu V$ after additional 300\,ms and subsequently faded within 200\,ms.
To introduce variability across trials, we randomly varied the location ($\le$10\,mm radius), latency ($\le\pm$200\,ms) and peak amplitude ($\le\pm$20\,$\mu V$).

The stationary electrode noise was modeled as white noise at each electrode with a participant and electrode specific amplitude within 0.5 to 1.5 $\mu V$.

We modeled pops as single electrode sources.
The pop waveform was modeled as a step with an amplitude of 100 $\pm$ 10 $\mu V$ and a subsequent exponential decay with a time-constant of 0.25 $\pm$ 0.08 $s^{-1}$.
To allow multiple pops per trial, we used 10 electrode pop sources that were iid.		
For each trial every pop source could get active with a 2\,\% probability at any electrode and time point within the interval [5, 10]\,s.

Electrode drifts were modeled as transient pink noise that was limited to the $[0.1,\,0.3]$\,Hz band.
The band-limted pink noise was weighted by a Tukey window so that the transient drifts were limited to the interval [3, 12]\,s.
Similar to the pops, we used 10 iid drift sources.		
For each trial every drift source would get active with a 2\,\% probability at one of the 64 electrodes.

We added the contribution of each source to the electrodes and stored the result in an EEGLAB dataset \cite{Delorme2004}.
For the reach task, we also saved the clean EEG signals $ \vec{X}_{clean} $. I.e., the EEG signals without contributions from the noise sources (electrode noise, pops and drifts).
The simulated dataset and the code to generate it are publicly available \cite{Kobler2019a}.

We evaluated the algorithms during the interval [5, 10]\,s by computing the signal to noise ratio (SNR) defined as 

\begin{equation}
	SNR(\vec{M}) = \frac{ ||\vec{X}_{clean} \circ \vec{M}||_2}{||\vec{X}_{clean} \circ \vec{M} - \vec{X}_{corrected} \circ \vec{M}||_2}
	\label{eq:eval:snr}
\end{equation}

with $\vec{X}_{clean}$ and $\vec{X}_{corrected}$ being $N_{electrodes}~\times~ N_{samples}~\times~ N_{trials}$ arrays of clean and corrected EEG signals.
We computed the SNR for PD artifact contaminated or non-artifact contaminated data by applying a $N_{electrodes}~\times~ N_{samples}~\times~ N_{trials}$ mask $\vec{M}$.
In case of PD artifact contaminated data, $\vec{M}$ indicated whether an array element was contaminated by a PD artifact.
Applying $\vec{M}$ on $\vec{X}$ extracted a vector of all PD artifact contaminated elements.
Complementary to the SNR, we estimated the MRCP by averaging the clean and corrected EEG signals over trials.

%\begin{outline}
%
%Evaluation paragraph:
%\begin{itemize}
%	\item We defined the signal to noise ratio (SNR) as 
%	$$ SNR(\vec{M}) = \frac{ ||\vec{X}_{clean} \circ \vec{M}||_2}{||\vec{X}_{clean} \circ \vec{M} - \vec{X}_{corrected} \circ \vec{M}||_2}  $$
%	with $\vec{X}_{clean}$ and $\vec{X}_{corrected}$ being $N_{electrodes}~\times~ N_{samples}~\times~ N_{trials}$ arrays of clean and corrected EEG signals.
%	We computed the SNR for PD artifact contaminated or non-artifact contaminated data by applying a $N_{electrodes}~\times~ N_{samples}~\times~ N_{trials}$ mask $\vec{M}$.
%	In case of PD artifact contaminated data, $\vec{M}$ coded whether an element was contaminated by a PD artifact.
%	Applying $\vec{M}$ on $\vec{X}$ extracted a vector of all PD artifact contaminated elements.
%	
%	\item Complementary to the SNR, we computed the MRCP by averaging the clean and corrected EEG signals over trials.
%\end{itemize}
%
%\end{outline}

\subsection{Real-world EEG dataset}

In addition to the simulation, we evaluated the algorithms on one real-world dataset.
%The dataset consisted of EEG signals acquired during a visually guided center-out reaching task in four directions.
The dataset consists of EEG recordings of 15 participants, while they performed visuomotor and oculomotor tasks.
The experimental conditions, tasks and equipment is described in detail in \cite{Kobler2018}.
Here, we analyzed the EEG signals during a center-out reaching task.
The participants were asked to make a center-out movement with a cursor after a target moved to a specific direction.
They operated the cursor by moving their right hand on a 2D surface.
In each trial, the target started to move at 2.5\,s and stopped at one of four possible positions at 3.0\,s.
The grand-average cursor movement onset was at 3.2\,s.

As in \cite{Kobler2018}, we pre-processed the EEG by resampling the signals of 64 EEG electrodes at 200\,Hz, applying a high-pass filter (0.25\,Hz cut-off frequency, Butterworth filter, eighth order, zero-phase), a band-stop filter (49 and 51\,Hz cut-off frequencies, Butterworth filter, fourth order, zero-phase), spherically interpolating bad channels, correcting eye artifacts and re-referencing to the common average reference (CAR).

The parameters of HEAR and ASR were calibrated to resting data, that were recorded at the beginning of the experiment according to the paradigm outlined in \cite{Kobler2017}.
To ensure clean calibration data, we applied automatic trial rejection criteria.
In detail, resting trials were rejected, if the EEG signal of any electrode exceeded a threshold ($\pm$200\,$\mu V$) or had a abnormal probability/variance/kurtosis ($\ge$ (6/4/6) standard-deviations beyond the mean).

We used two criteria to evaluate the correction algorithms.
First, as for the simulated data, we computed the MRCP through averaging the trials during the center-out reaching task.
Second, we applied the above defined automatic outlier trial detection criteria (threshold/probability/variance/kurtosis) to the uncorrected and corrected EEG signals.
Assuming that the correction algorithms improve the SNR, fewer trials should be marked as outliers for the corrected EEG signals.

\section{Results}

We present grand-average results which are summarized by the mean across participants.
Variability over participants is summarized by the standard-error of the mean, if not specified otherwise.

\begin{figure}[thpb]
	\centering
	\includegraphics{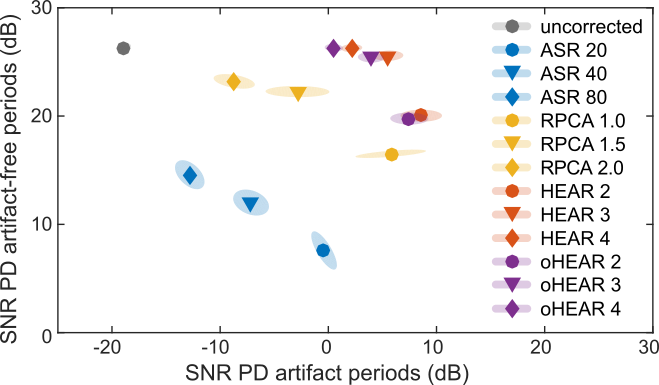}
	\caption{
		SNR during PD artifact and PD artifact-free periods for the simulated EEG dataset.
		Each marker represents the mean across the 15 participants, while the shaded area summarizes the scatter over participants (matrix square root of the covariance matrix).
		The color indicates the algorithm and the marker the value of the sensitivity parameter ($k$ for ASR, $\lambda_0$ for RPCA, and $\phi$ for (o)HEAR). For example, a blue circle identifies the result for ASR with $k = 20$.
	}
	\label{fig:sim:snr}
\end{figure}

In the case of the simulated dataset, we had access to all sources.
This allowed us to compute the SNR according to (\ref{eq:eval:snr}).
Figure\,\ref{fig:sim:snr} depicts the SNR during PD artifact and PD artifact-free periods before and after the correction algorithms were applied.
The SNR of the uncorrected signals was -19$\pm$0.2\,dB and 26.3$\pm$0.1\,dB during PD artifact and PD artifact-free periods, respectively.
Both SNR metrics are informative.
If an algorithm overcorrects the signals, the SNR during artifact-free periods is low.
If an algorithm undercorrects the PD artifacts, the SNR during artifact periods is low.
The ideal correction algorithm would maximize the SNR during both periods.
%It is apparent that the three types of algorithms operated in different regimes.

Regarding PD artifact periods, all algorithms increased the SNR compared to the uncorrected EEG.
The increase depended strongly on the sensitivity parameters of the algorithms.
%For largest sensitivity, the SNR was 0.4\,dB for ASR, 5.7\,dB for RPCA, 8.6\,dB for HEAR and 7.5\,dB for oHEAR.
A higher sensitivity lead to higher SNR during artifact periods at the cost of lower SNR during non-artifact periods.
%For increased sensitivity had also considerable influence on the SNR during artifact free periods.
%This dependence can be approximated with a line with negative slope for all algorithms.
For example, ASR (blue color) corrected the PD artifacts almost as good as RPCA (yellow color), but removed more activity during PD artifact-free periods for comparable sensitivities.
The best trade-off between over- and under-correction was achieved by (o)HEAR with $\phi = 3$, RPCA with $\lambda_0 = 1.5$ and ASR with $k = 40$.

\begin{figure}[thpb]
	\vspace{6pt}
	\centering
	\includegraphics{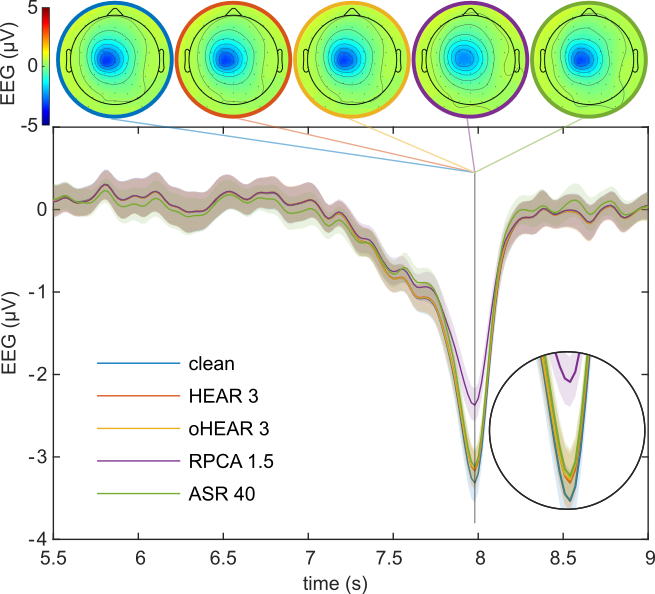}
	\caption{
		Grand-average MRCP at electrode C1 for the simulated EEG dataset.
		The topographic distribution of the potential at the peak negativity is summarized in the top row.
		The outline color indicates the algorithm.
		A 100\,ms triangular window (zero-phase) was used to smooth the signals before they were plotted.
		The inset shows a closer view of the differences at peak negativity.
	}
	\label{fig:sim:mrcp}
\end{figure}

We were not only interested in the impact of the correction algorithms on continuous EEG signals, but also on the simulated MRCP.
The grand-average MRCP is displayed in Figure\,\ref{fig:sim:mrcp}.
The topographic plots at the peak negativity demonstrate that all algorithms preserved the MRCP waveform.
Compared to the MRCP of the clean data (blue contour), RPCA attenuated the peak most (violet contour), with a reduction in peak amplitude of approximately 1\,$\mu V$.
The other algorithms attenuated the peak only negligibly ($\le 0.2\,\mu V$).

\begin{figure}[thpb]
		\vspace{6pt}		
		\centering
		\includegraphics{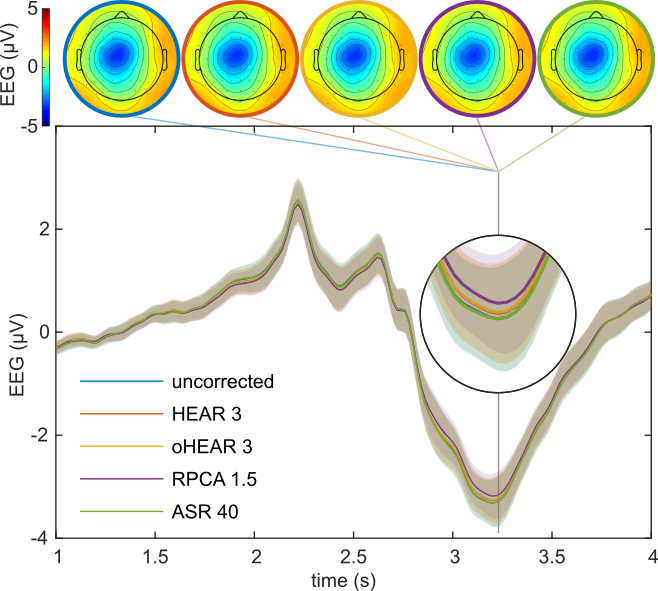}
		\caption{
			Grand-average MRCP at electrode C1 for the real EEG dataset.
			The topographic plots in the top summarize the distribution of the potential at the peak negativity.
			The outline color indicates the algorithm.
			We used a 100\,ms triangular window (zero-phase) to smooth the signals before they were plotted.
			The inset shows a closer view of the differences at peak negativity.
		}
		\label{fig:real:mrcp}
	\end{figure}

\begin{figure}[thpb]
		\centering
		\includegraphics{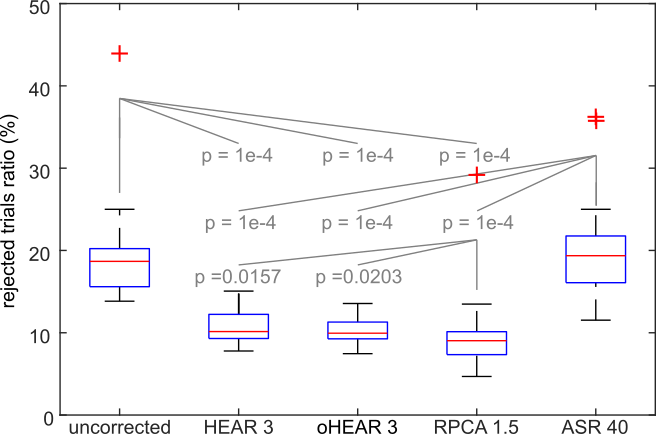}
		\caption{
	    	Boxplots summarizing the fraction of trials marked as outliers across the participants of the real EEG dataset.
	    	Trials were automatically marked as outliers according to threshold/probability/variance/kurtosis criteria.
	    	Two-sided, paired Wilcoxon sign-rank tests were used to detect significant differences between the algorithms.
	    	We controlled the false discovery rate at significance level $\alpha = 0.05$ for 10 tests ($p_{crit} = 0.0203$) \cite{Benjamini1995}.
   		}
		\label{fig:real:rej}
	\end{figure}

Figure\,\ref{fig:real:mrcp} displays the grand-average MRCP for the real EEG dataset.
All algorithms preserved a clear MRCP waveform.
We only observed negligible differences ($\le 0.2\,\mu V$) compared to the uncorrected EEG.
The algorithm specific averages were computed after outlier trials were automatically detected and discarded according to the threshold/probability/variance/kurtosis criteria.
The fraction of trials that were marked as outliers is displayed in Figure\,\ref{fig:real:rej}.
In the case of uncorrected EEG, a median of 18.7\,\% of the trials were marked as outliers.
The result did not significantly differ for ASR, while for HEAR, oHEAR and RPCA significantly fewer trials were marked as outliers.
The fraction of outlier trials did not significantly differ between HEAR (10.1\,\%) and oHEAR (9.9\,\%).
Compared to HEAR and oHEAR, RPCA could sightly, yet significantly reduce the fraction to 9.0\,\%.

%\begin{table}[h]
%\caption{An Example of a Table}
%\label{table_example}
%\begin{center}
%\begin{tabular}{|c||c|}
%\hline
%One & Two\\
%\hline
%Three & Four\\
%\hline
%\end{tabular}
%\end{center}
%\end{table}

\section{DISCUSSION}

PD artifacts can significantly reduce the number of available trials in an offline analysis, and deteriorate closed-loop BCI control.
We proposed HEAR - a simple, yet efficient algorithm to correct transient PD artifacts.
The presented simulation results show that HEAR and oHEAR improved the SNR during PD artifacts by approximately 25\,dB, and at the same time maintained a high SNR during PD artifact-free periods.
State of the art offline (RPCA) and online (ASR) correction algorithms were clearly inferior to (o)HEAR.
Compared to uncorrected data, the application of (o)HEAR to real-world EEG signals resulted in a significantly reduced fraction of outlier trials, while slow potentials such as MRCPs could be preserved.

Regarding RPCA, the simulated and real EEG dataset results were not entirely consistent.
While RPCA attenuated the simulated MRCP peak considerably stronger than the other algorithms, the difference to the other algorithms was only negligible for the real EEG dataset.
The SNR performance of RPCA (Figure\,\ref{fig:sim:snr}) was clearly lower compared to HEAR for the simulated data.
However, RPCA marginally (1.1\,\%) but significantly reduced the fraction of outlier trials (Figure\,\ref{fig:real:rej}) compared to HEAR for the real EEG dataset.
%The difference in the fraction of outlier trials was marginal (1.0\,\%).
%The fraction of outlier trials is an indirect indicator of the SNR and has to be interpreted with caution.
Taken together, it is difficult to generalize the influence of RPCA on the desired brain activity across datasets.

In case of ASR, the SNR (Figure\,\ref{fig:sim:snr}) could be only improved at the cost of lower SNR druing PD artifact-free periods.
Also, the fraction of outlier trials (Figure\,\ref{fig:real:rej}) could not be significantly reduced by ASR.
Since ASR is based on PCA on short time-windows (0.5\,s), and PCA is sensitive to drifts, we think that a longer time-window could have improved the correction quality.
However, we decided to use the default window size because ASR introduces a processing delay of 0.5 times the window size to achieve the best online correction quality.
%The processing delay that ASR introduces to a pipeline in an online scenario can be up to 1.5 times the window size.
%If the pipeline processes the data already in chunks of 0.5\,s, this reduces to 0.5 times the window size.
For the considered window size, ASR introduces a delay of 250\,ms. In case of neuromodulation studies, adding more delay in the detection of transient events such as the onset of movement can be critical \cite{MrachaczKersting2012}.
(o)HEAR uses an exponential smoothing filter which belongs to the class of infinite impulse response filters.
As such, the processing delay is frequency specific.
For oHEAR with $t_{est} = 0.25\,s$ it is $\le 105\,$ms and declines with rising frequencies.

The correction quality of (o)HEAR depends on the spatial resolution of the electrodes.
The higher the spatial resolution (i.e., number of electrodes), the better is the interpolation quality.
%However, from EEG source imaging we know that the localization precision increases drastically until the number of equally spaced electrodes reaches 64 and then begins to saturate \cite{Michel2004}.
%One has also to keep in mind that the probability that any PD artifact happens increases with the number of electrodes.
The presented results demonstrate that 64 equally spaced electrodes were sufficient to outperformed state of the art methods.
The interpolation quality could be further improved by using spherical splines instead of Euclidean distances \cite{Michel2004,Perrin1989}.

Online, not only the correction quality, but also the computational complexity matters.
For a given electrode configuration, the interpolation matrix $\vec{D}$ can be pre-computed for $k$ nearest neighbors ($k$NN).
Then, the matrix multiplications in (\ref{eq:hear:correction}) simplify to $k$ element-wise multiplications, which can be computed in $O(k \cdot N_{channels})$.
This is considerably faster compared to ASR whose run-time is mainly constrained by PCA which can be computed in $O(max\{N_{samples}\cdot N_{channels}^2, N_{channels}^3\})$.

We designed (o)HEAR to remove PD artifacts, which are typically active at single electrodes.
Other types of transient, high-variance artifacts such as EMG and sweat artifacts typically do not meet this assumption.
In that case, the correction quality of (o)HEAR is certainly going to deteriorate.
Still, one can compute the probability that a transient, high-variance artifact cannot be corrected by (o)HEAR. If $\vec{D}$ is applied on $\vec{P}[n]$, the result is a vector of probabilities that indicate how likely each $k$NN estimate is contaminated by an artifact.
If a threshold is applied, EMG and sweat artifacts can be detected.

%\section{CONCLUSIONS}
%
%\begin{todoe}
%	Outline!
%	\begin{itemize}
%		\item ASR not suitable for low-frequency artifacts. If both low-frequency amplitude and band-power features in higher frequency bands are used in a BCI. ASR could be used to reduce EMG artifacts and HEAR to correct PD artifacts.
%	\end{itemize}
%\end{todoe}

%A conclusion section is not required. Although a conclusion may review the main points of the paper, do not replicate the abstract as the conclusion. A conclusion might elaborate on the importance of the work or suggest applications and extensions. 

\addtolength{\textheight}{-3cm}   % This command serves to balance the column lengths

%%%%%%%%%%%%%%%%%%%%%%%%%%%%%%%%%%%%%%%%%%%%%%%%%%%%%%%%%%%%%%%%%%%%%%%%%%%%%%%%

%%%%%%%%%%%%%%%%%%%%%%%%%%%%%%%%%%%%%%%%%%%%%%%%%%%%%%%%%%%%%%%%%%%%%%%%%%%%%%%%

%%%%%%%%%%%%%%%%%%%%%%%%%%%%%%%%%%%%%%%%%%%%%%%%%%%%%%%%%%%%%%%%%%%%%%%%%%%%%%%%
%\section*{APPENDIX}

\section*{ACKNOWLEDGMENT}

The authors acknowledge Joana Pereira, Catarina Lopes Dias and Lea Hehenberger for their valuable comments.
This work has received funding from the European Research Council (ERC) under the European Union's Horizon 2020 research and innovation programme (Consolidator Grant 681231 'Feel Your Reach').
V. M. received additional funding from a Marco Polo scholarship sponsored by the University of Bologna.

\section*{CODE \& DATA AVAILABILITY}

We encourage a widespread use of (o)HEAR and, therefore, provide an open-source reference implementation of (o)HEAR (\url{https://bci.tugraz.at/research/software/#c218405}).
To support future improvements and ease comparability of algorithms, we also provide the dataset of simulated EEG signals and the code to generate it \cite{Kobler2019a}.

%%%%%%%%%%%%%%%%%%%%%%%%%%%%%%%%%%%%%%%%%%%%%%%%%%%%%%%%%%%%%%%%%%%%%%%%%%%%%%%%

\section*{REFERENCES}

\printbibliography[heading=none]

\end{document}